\documentclass[aps,prd,twocolumn,superscriptaddress,floatfix,nofootinbib]{revtex4-2}

\usepackage{amsmath,amssymb,amsthm,mathtools,bm}
\usepackage{graphicx}
\usepackage{hyperref}
\usepackage{microtype}

\hypersetup{colorlinks=true,linkcolor=blue,citecolor=blue,urlcolor=blue}
\allowdisplaybreaks[4]

\newcommand{\id}{\mathbf 1}
\newcommand{\cV}{\mathcal V}
\newcommand{\dd}{\mathrm d}
\newcommand{\Spec}{\operatorname{Spec}}
\newcommand{\Boxop}{\Box}
\newcommand{\Tr}{\operatorname{Tr}}
\newcommand{\NO}[1]{\mathopen{:} #1 \mathclose{:}}
\newcommand{\sA}{\mathsf A}
\newcommand{\sB}{\mathsf B}
\newcommand{\sC}{\mathsf C}
\newcommand{\sD}{\mathsf D}
\newcommand{\st}{\mathsf t}
\newcommand{\Up}{|{\Uparrow}\rangle}
\newcommand{\cW}{\mathcal W}
\newcommand{\cG}{\mathcal G}
\newcommand{\cS}{\bm{\mathcal S}}
\newcommand{\fr}{\mathfrak f}
\newcommand{\CRL}{\mathsf C_{RL}}
\newcommand{\qconst}{\mathcal Q}

\begin{document}

\title{Integrable Chiral Quantum Field Theories with Spacetime-Dependent Interactions}

\author{Pradip Kattel}
\email{pradip.kattel@unige.ch}
\affiliation{Department of Quantum Matter Physics, University of Geneva, Quai Ernest-Ansermet 24, 1211 Geneva, Switzerland}

\begin{abstract}
We construct a class of integrable chiral fermionic quantum field theories with interactions that depend on both space and time. The construction starts from a unitary, regular, difference-form two-body scattering matrix satisfying braiding unitarity and the Yang--Baxter equation. Each freely moving right- or left-moving particle carries a characteristic coordinate that remains constant along its trajectory, and the scattering matrix depends on differences of these coordinates. The many-body amplitudes form a discrete connection over the coordinate-ordering sectors. The Yang--Baxter equation gives local path independence, while periodic winding on a spatial circle produces quantum Knizhnik--Zamolodchikov (qKZ) difference equations expressing global flatness. Once a reference amplitude satisfying the exchange and qKZ constraints is specified, these transport equations determine the exact wavefunction in every fixed-particle-number sector. Independent reparametrizations of the right- and left-moving characteristics give couplings that depend on both space and time. For the chiral $SU(2)$ Gross--Neveu realization, we solve the qKZ equations for this full characteristic family using off-shell Bethe vectors. The resulting amplitudes are written in the characteristic labels, and their pullbacks give exact many-body wavefunctions for the corresponding space-time-dependent couplings. The two-body amplitude is a ratio of Gamma functions that resums the phase shifts from successive scattering processes. Equal-slope affine maps give a spatially homogeneous specialization in which the coupling depends only on time, and the Bethe ansatz with constant coupling is recovered when the qKZ step vanishes.
\end{abstract}

\maketitle

\section{Introduction}\label{sec:intro}

For an autonomous quantum system, time-translation symmetry is generated by a time-independent Hamiltonian. Integrability is characterized by a commuting family of conserved quantities that includes this generator. When the Hamiltonian depends explicitly on time, time-translation invariance is lost, and the evolution is generated by a time-ordered family of Hamiltonians. Even if the Hamiltonian at each fixed time has commuting conserved quantities, these quantities may depend on time. Their compatibility with the time-ordered evolution becomes an additional integrability condition. The scattering or Lax structure must be compatible with the time evolution.

For partial differential equations, this type of consistency is often formulated through an overdetermined auxiliary linear system. Compatibility of the auxiliary equations requires the associated connection to be flat, which in local coordinates is expressed as a zero-curvature condition~\cite{ZakharovShabat1974,AblowitzKaupNewellSegur,DrinfeldSokolov1985,Dickey2003}. We shall use the same compatibility principle for a chiral fermionic quantum field theory. The many-body amplitudes are assigned to coordinate-ordering sectors and related between neighboring sectors by two-body scattering matrices. These transport maps form a discrete connection. The Yang--Baxter equation gives local path independence, while periodic winding on a spatial circle gives quantum Knizhnik--Zamolodchikov difference equations whose compatibility expresses global path independence.

Similar conditions appear in exactly solvable time-dependent quantum systems and integrable field theories~\cite{Sinitsyn2018,Yuzbashyan2018,KomatsuSakamotoWallbergYamazaki2026}. Recent chiral constructions have produced explicit time-dependent couplings $g(t)$, exact many-body wavefunctions, and qKZ formulations for the Kondo and $SU(2)$ Gross--Neveu models~\cite{PasnooriKondo2025,PasnooriYuzbashyan2025,Pasnoori2026,PasnooriGN2026}. Ref.~\cite{Kattel2026} used this approach for interacting $1+1$-dimensional chiral quantum field theories by transporting the spectral parameters of an autonomous unitary difference-form scattering matrix along the free right- and left-moving characteristics. Spatial homogeneity was imposed there, so that the right-left scattering data, and hence the coupling obtained from the inverse Cayley map, depended only on time. This requires affine characteristic maps with equal slopes and gives a coupling $g(t)$. Here we allow the coupling to depend on space as well as time. Independent nonaffine reparametrizations of the two characteristics preserve the factorized transport structure while producing integrable couplings $g(x,t)$ with genuine space-time dependence.

We also solve the transport equations explicitly. Once the characteristic labels and the two-body matrices are fixed, the only remaining input is a reference amplitude satisfying the exchange and qKZ constraints. For the rational $SU(2)$ realization we construct this amplitude in closed form in every fixed-particle-number and total-spin sector. We use the off-shell Bethe ansatz~\cite{Babujian1993,BabujianFlume1994,BabujianKarowskiZapletal1997}, in which auxiliary Bethe roots are summed over a lattice instead of being constrained by Bethe equations, so that every amplitude is a Jackson integral with an explicit Gamma-function weight~\cite{Matsuo1993,ReshetikhinJackson1992,TarasovVarchenko1994}. The periodic quasi-constants of the qKZ system specify the free chiral data on the ring. At real coupling, the Jackson integrals lie on the boundary of absolute convergence. Since the coupling decays as the inverse characteristic separation, successive scattering processes produce a logarithmic phase. We define the sums by the $\eta\to\eta+i0$ prescription, as discussed below in connection with the Coulomb phase~\cite{Dollard1964}. Right- and left-moving clusters decouple at late times up to a logarithmic phase with a spin-dependent charge. In the zero-step limit, the qKZ difference equations reduce to the commuting transfer-matrix eigenvalue problem of the constant-coupling chiral-invariant Gross--Neveu model, recovering the Bethe-ansatz diagonalization of Refs.~\cite{AndreiLowenstein1979,DestriLowenstein1982}. In a separate semiclassical limit, they yield the Bethe equations of the inhomogeneous XXX chain, with the characteristic labels serving as inhomogeneities.

The paper is organized as follows. Section~\ref{sec:characteristics} introduces the characteristic labels and the scattering variable. Section~\ref{sec:transport} formulates the factorized transport over coordinate-ordering sectors, and Sec.~\ref{sec:qkz} derives the qKZ connection from winding on the circle, identifies its quasi-constants with free chiral data, and gives the holonomy form of the initial-value problem. Section~\ref{sec:GN} presents the chiral $SU(2)$ Gross--Neveu realization. Section~\ref{sec:solution} contains the exact solution of the qKZ transport, and Sec.~\ref{sec:physics} discusses its physical consequences. Section~\ref{sec:conclusions} concludes. Appendix~\ref{app:proof} proves the Jackson-integral representation.

\section{Characteristic labels and the scattering variable}\label{sec:characteristics}

We shall consider a chiral fermionic quantum field theory with an internal one-particle space $\cV$,
\begin{equation}
 i\partial_t|\Psi(t)\rangle=H_{\rm ch}(t)|\Psi(t)\rangle , \label{eq:Sch}
\end{equation}
where the explicit time dependence is carried by a local four-fermion interaction.

We shall first formulate the construction abstractly in a representation-invariant way. Let $S(u)$ be a unitary, regular, difference-form two-body scattering matrix satisfying braiding unitarity and the Yang--Baxter equation. The argument $u$ is built from conserved labels carried by the free chiral characteristics. These labels are constant along free trajectories and enter the two-body scattering matrices. They are not rapidities or momenta. For a right and a left mover, we write
\begin{equation}
 \xi_R(x,t)=f_R(x-t),\qquad \xi_L(x,t)=f_L(x+t),
\label{eq:chars}
\end{equation}
where $f_R$ and $f_L$ are twice continuously differentiable real-valued functions satisfying $f_R'(s)>0$ and $f_L'(s)>0$. Thus,
\begin{equation}
 (\partial_t+\partial_x)\xi_R=0,\qquad (\partial_t-\partial_x)\xi_L=0 .
\label{eq:charPDE}
\end{equation}
The labels are constant along the corresponding free trajectories. Since $f_R'(s)>0$ and $f_L'(s)>0$, both maps are increasing and one-to-one. They preserve the ordering of trajectories within each chirality and ensure that equal labels correspond to equal positions. The functions $f_R$ and $f_L$ relabel the characteristics without changing the free trajectories or the scattering construction. When the scattering data depend on the label difference, nonaffine choices make the resulting interaction depend on both $x$ and $t$ in the laboratory frame~\cite{Kattel2026}.

Introduce the null coordinates $a=x-t$ and $b=x+t$. The map
\begin{equation}
 (x,t)\longmapsto(a,b)=(x-t,x+t)
\label{eq:nullmap}
\end{equation}
has inverse $x=(a+b)/2$ and $t=(b-a)/2$. Free right movers keep $a$ fixed, while free left movers keep $b$ fixed. The characteristic labels are the independent reparametrizations $\xi_R=f_R(a)$ and $\xi_L=f_L(b)$. The two-body scattering matrix depends on the difference between these labels.

The scattering variable is the difference
\begin{equation}
 u(x,t)=\xi_L-\xi_R =f_L(x+t)-f_R(x-t).
\label{eq:u}
\end{equation}
Since $u$ is a difference of functions of the two null coordinates,
\begin{equation}
\Boxop u=0,\qquad \Boxop=\partial_t^2-\partial_x^2.
\label{eq:boxu}
\end{equation}
This equation holds for $u$ independently of the Hamiltonian coupling. The relation between $u$ and the coupling is model dependent. Once a contact prescription is fixed, the Cayley map determines a function $\Upsilon$ such that
\begin{equation}
 u=\Upsilon\left(g(x,t)\right).
\label{eq:upsilon}
\end{equation}
As in Ref.~\cite{Kattel2026}, this relation is fixed by the contact prescription together with the choice of scattering matrix. Only after specifying this model-dependent map does the geometric condition become
\begin{equation}
 \Boxop\Upsilon\left(g(x,t)\right)=0 .
\label{eq:upsilonwave}
\end{equation}
Conversely, on a null rectangle, every twice continuously differentiable solution of $\Boxop u=0$ has the local form
\begin{equation}
 u=\mathcal A(x-t)+\mathcal B(x+t).
\label{eq:dAlembert}
\end{equation}

This is the local d'Alembert form of the characteristic scattering variable.

\section{Factorized transport over ordering sectors}\label{sec:transport}

For three labels, the two possible sequences of pairwise exchanges must produce the same amplitude. Their local path independence is expressed by the Yang--Baxter equation
\begin{equation}
\begin{split}
&S_{12}(\xi_1-\xi_2)S_{13}(\xi_1-\xi_3)S_{23}(\xi_2-\xi_3)\\
&\qquad =S_{23}(\xi_2-\xi_3)S_{13}(\xi_1-\xi_3)S_{12}(\xi_1-\xi_2).
\end{split}
\label{eq:YBE}
\end{equation}
This ensures that the amplitude does not depend on the order of the exchanges~\cite{Yang1967,Yang1968,KorepinBook,SmirnovBook}.

For particle $j$ with chirality $\chi_j=\pm1$, define
\begin{equation}
 \xi_j=f_{\chi_j}(x_j-\chi_j t).
\label{eq:manylabels}
\end{equation}
Let $Q$ denote a coordinate-ordering region, $x_{Q_1}<\cdots<x_{Q_N}$. In that region, set
\begin{equation}
 \Psi_Q(\bm x,t)=F_Q(\xi_1,\ldots,\xi_N).
\label{eq:pullback}
\end{equation}
Since the labels are constant along the free trajectories,
\begin{equation}
 i\partial_t\Psi_Q =-i\sum_j\chi_j\partial_{x_j}\Psi_Q
\label{eq:freeN}
\end{equation}
inside each ordering region.

Neighboring regions differ by an adjacent exchange $s_a$, which swaps the entries $a$ and $a+1$ of $Q$, and their amplitudes are related by
\begin{equation}
 F_{Qs_a}(\bm\xi)= S_{Q_aQ_{a+1}} \left(\xi_{Q_a}-\xi_{Q_{a+1}}\right)F_Q(\bm\xi).
\label{eq:exchange}
\end{equation}
For opposite chiralities, this is the physical contact-scattering condition. For equal chiralities, the two particles have the same velocity and do not undergo a physical contact scattering event. Their amplitudes are nevertheless related by the same auxiliary exchange matrix $S$, so that all pairwise transports belong to one Yang--Baxter algebra. Because equal-chirality particles move parallel to their coincidence hyperplane, the free operator in Eq.~\eqref{eq:freeN} differentiates only along that hyperplane, and the auxiliary transport across it generates no contact term.

Let $\tau_{a,a+1}$ exchange the two spatial arguments while $P_{a,a+1}$ exchanges their internal labels. Fermionic statistics requires the amplitude to be odd under the simultaneous exchange, $P_{a,a+1}\tau_{a,a+1}\Psi=-\Psi$. For two particles of equal chirality, exchanging their spatial arguments exchanges their labels and maps the ordering region $Q$ onto $Qs_a$. Combined with Eq.~\eqref{eq:exchange}, the antisymmetry condition becomes a constraint on the label dependence of the amplitudes. For adjacent particles $a,a+1$ of equal chirality in the reference ordering, it reads
\begin{equation}
\begin{aligned}
 F(\ldots,\xi_{a+1},\xi_a,\ldots)={}&-P_{a,a+1}S_{a,a+1}(\xi_a-\xi_{a+1})\\
 &\times F(\ldots,\xi_a,\xi_{a+1},\ldots).
\end{aligned}
\label{eq:stat}
\end{equation}
Because $f_R$ and $f_L$ are one-to-one, equal labels for equal-chirality particles occur only when they occupy the same position. There $S(0)=P$, so the auxiliary transport across the coincidence hyperplane is the bare permutation of internal labels, and Eq.~\eqref{eq:stat} is the antisymmetry condition $P_{a,a+1}\tau_{a,a+1}\Psi=-\Psi$ on that hyperplane. Since equal-chirality particles do not interact, the amplitude is not required to be continuous across it. Away from coincidence, Eq.~\eqref{eq:stat} supplies the auxiliary exchange transport required by the Yang--Baxter construction. The Yang--Baxter equation and braiding unitarity then make the amplitudes independent of the sequence of exchanges used to reach a given region.

For fixed chirality content, with $N_R+N_L=N$, the resulting state is
\begin{equation}
\begin{aligned}
 |\Psi_{N_R,N_L}(t)\rangle &=\frac{1}{N_R!N_L!}\sum_{Q\in S_N} \int_{\mathcal C_Q}\dd^N x\, F_Q^{a_1\cdots a_N}(\bm\xi(\bm x,t))\\
 &\quad\times \psi_{\chi_1a_1}^\dagger(x_1)\cdots \psi_{\chi_Na_N}^\dagger(x_N)|0\rangle .
\end{aligned}
\label{eq:fixedNstate}
\end{equation}
Here $\mathcal C_Q$ is the region $x_{Q_1}<\cdots<x_{Q_N}$. We label the particles so that $j=1,\ldots,N_R$ are right movers and $j=N_R+1,\ldots,N$ are left movers, and we fix the reference region $Q_0$ to be the identity ordering $x_1<\cdots<x_N$, in which every right-left pair is in its incoming configuration. We write $F\equiv F_{Q_0}$. If $Q=Q_0s_{a_1}\cdots s_{a_m}$, then
\begin{equation}
 F_Q=\mathcal U_Q(\bm\xi)F,\qquad \mathcal U_Q= S_{a_m}(\bm\xi)\cdots S_{a_1}(\bm\xi),
\label{eq:referenceamplitude}
\end{equation}
with the ordered labels inserted according to Eq.~\eqref{eq:exchange}. The reference amplitude is constrained by the fermionic exchange relation in Eq.~\eqref{eq:stat} and, on the circle, by the qKZ equations below.

\section{Winding on the circle and the qKZ connection}\label{sec:qkz}

We impose periodic boundary conditions on a ring of circumference $\ell$, identifying $x\sim x+\ell$, and work on its universal cover. A winding around the ring is represented there by $x\to x+\ell$.

Compatibility with winding requires the characteristic maps to be quasiperiodic lifts,
\begin{equation}
f_R(s+\ell)=f_R(s)+\varpi,\qquad f_L(s+\ell)=f_L(s)+\varpi ,
\label{eq:quasi}
\end{equation}
with $\varpi>0$. The functions are defined on the unwrapped line and increase by $\varpi$ after one circuit of length $\ell$.

The common shift cancels in the difference $\xi_L-\xi_R$, so the interaction is periodic in space. To construct the qKZ equation, we take particle $j$ to be a right mover and carry it once around the circle in the positive $x$ direction. For this circuit,
\begin{equation}
\xi_j(x_j+\ell,t)=\xi_j(x_j,t)+\varpi .
\label{eq:winding}
\end{equation}
This is the lift relation for the right-moving characteristic along the circuit used below. Iterating it gives $f_R(s+n\ell)=f_R(s)+n\varpi$, so repeated right-moving windings shift the corresponding label by integer multiples of $\varpi$. These shifts are the steps of the qKZ equations.

Write $\Delta_{jk}=\xi_j-\xi_k$. Let particle $j$ move to the right through particles $j+1,\ldots,N$. Each passage multiplies the amplitude by $S_{jk}$ evaluated at the current label difference, Eq.~\eqref{eq:exchange}. At $x=\ell$ the particle reappears at $x=0$, where by Eq.~\eqref{eq:winding} its label is $\xi_j-\varpi$, and it then passes particles $1,\ldots,j-1$ before returning to its reference position. Single-valuedness on the ring requires
\begin{equation}
 F(\bm\xi-\varpi e_j) =\mathcal K_j(\bm\xi;\varpi)F(\bm\xi),
\label{eq:qKZ}
\end{equation}
where $e_j$ is the unit vector along the $j$th label and $\mathcal K_j$ is the product of the two-body matrices met on one circuit,
\begin{equation}
\begin{split}
 \mathcal K_j(\bm\xi;\varpi)=& S_{j,j-1}(\Delta_{j,j-1}-\varpi)\cdots S_{j1}(\Delta_{j1}-\varpi)\\
 &\times S_{jN}(\Delta_{jN})\cdots S_{j,j+1}(\Delta_{j,j+1}).
\end{split}
\label{eq:Kj}
\end{equation}
Winding two right movers in the $j$ and $i$ directions in the two possible orders leads to the same final label configuration. The transport operators must agree,
\begin{equation}
\begin{aligned}
 \mathcal K_i(\bm\xi-\varpi e_j;\varpi)\mathcal K_j(\bm\xi;\varpi)
 =&\mathcal K_j(\bm\xi-\varpi e_i;\varpi)\\
 &\times\mathcal K_i(\bm\xi;\varpi).
\end{aligned}
\label{eq:qKZflat}
\end{equation}
This is the global flatness condition for winding transport. The Yang--Baxter equation and braiding unitarity of the two-body matrices ensure this compatibility, while Eq.~\eqref{eq:qKZ} imposes single-valuedness on the circle. Once a solution of the qKZ system is chosen for the reference amplitude, the exchange relations and the Yang--Baxter equation determine the amplitudes in every coordinate-ordering sector. The characteristic pullback in Eq.~\eqref{eq:pullback} then gives their complete space-time dependence. The qKZ equations are matrix difference equations in the characteristic labels~\cite{Smirnov1986,FrenkelReshetikhin1992,Cherednik1992,Reshetikhin1992}. Related time-dependent chiral $SU(2)$ theories have also been treated explicitly using qKZ and generalized Bethe-ansatz methods~\cite{Pasnoori2026,PasnooriGN2026}. 

\subsection{Quasi-constants and free chiral data}\label{sec:quasiconst}

The qKZ system is linear. If $F$ is one solution and $\qconst(\bm\xi)$ is any scalar function with period $\varpi$ in every label, then $\qconst F$ is another solution. These periodic functions are the freely chosen chiral data. They are functions of the winding coordinates
\begin{equation}
 z_j=e^{2\pi i\xi_j/\varpi}=\exp\!\left[\frac{2\pi i}{\varpi}f_{\chi_j}(x_j-\chi_jt)\right].
\label{eq:zvars}
\end{equation}
Because of Eq.~\eqref{eq:quasi}, each $z_j$ is a single-valued function on the ring, and because it depends on $x_j-\chi_jt$ only, it solves the free chiral equation. For equal-slope affine maps, $f_\chi(s)=\kappa s+\beta_\chi$, one has $\varpi=\kappa\ell$ and $z_j^n\propto e^{2\pi i n(x_j-\chi_jt)/\ell}$, a free mode of momentum $2\pi n/\ell$. For nonaffine maps the powers $z_j^n$ are warped free modes, still exact solutions of the free chiral equation on the ring. The quasi-constants specify the free chiral data; the qKZ solutions account for the scattering. The general reference amplitude is a combination of basis solutions with quasi-constant coefficients.

\subsection{Evolving the state from its initial data}\label{sec:holonomy}

The winding equations also give a direct way to evolve the state from its initial data, without first using the explicit integral formulas. Let $\mathfrak D=\prod_{j}[f_{\chi_j}(0),f_{\chi_j}(0)+\varpi)$. At $t=0$, every configuration on the ring corresponds to a label vector in $\mathfrak D$. Initial wavefunctions that obey the contact and exchange rules, Eqs.~\eqref{eq:exchange} and~\eqref{eq:stat}, determine $F$ throughout this starting region. Repeated right-moving windings then extend the result along the corresponding label directions,
\begin{equation}
 F(\bm\xi-\varpi\bm n)=\mathcal K^{(\bm n)}(\bm\xi)F(\bm\xi),\qquad \bm\xi\in\mathfrak D,\quad \bm n\in\mathbb Z_{\ge0}^N.
\label{eq:holonomy}
\end{equation}
Here $\mathcal K^{(\bm n)}$ is the ordered product of the matrices $\mathcal K_j$ encountered along a sequence of right-moving circuits. Each matrix is evaluated at the label vector at which its circuit begins. Equation~\eqref{eq:qKZflat} makes the result independent of the order in which the circuits are performed. Physically, every completed circuit produces another sequence of two-body scattering processes. Thus, the state at time $t$ is obtained from the state at $t=0$ by multiplying the scattering matrices for those processes, together with the auxiliary equal-chirality matrices that only relabel the ordering sectors. The number of right-left scattering processes in a pair grows as $2|t|/\ell$. The explicit solutions in Sec.~\ref{sec:solution} resum this scattering history in closed form; Eq.~\eqref{eq:prodrep} is the simplest example.

The explicit Hamiltonian realization below uses the bare chiral fermion fields and their bare Fock vacuum $|0\rangle$. We carry out the construction in every finite-particle-number sector above this vacuum. Dressed formulations will be discussed as a direction for subsequent work.

\section{Chiral $SU(2)$ Gross--Neveu realization}\label{sec:GN}

We consider a chiral $SU(2)$ Gross--Neveu realization~\cite{GrossNeveu1974}. Let $P$ permute the two factors of $\cV=\mathbb C^2$ and define $\Pi_\pm=(\id\pm P)/2$. The rational two-body scattering matrix is
\begin{equation}
 S(u)=\frac{u\id+i\eta P}{u+i\eta} =\Pi_++\frac{u-i\eta}{u+i\eta}\Pi_- ,\qquad \eta\in\mathbb R\setminus\{0\} .
\label{eq:S}
\end{equation}
The value $\eta=0$ gives the trivial free matrix $S(u)=\id$ and is excluded from the nontrivial realization considered below. The matrix obeys $S_{21}(u)=S_{12}(u)$, $S(u)S(-u)=\id$, and Eq.~\eqref{eq:YBE}.

Periodic boundary conditions impose one further consistency requirement. On the universal cover, winding a particle once around the circle changes its ordering relative to every other particle. The transport rules must apply to every pair of particles, including pairs with the same chirality. We use one common two-body matrix $S$ for all such exchanges.

The right- and left-moving labels vary independently because they are functions of the two independent null coordinates $x-t$ and $x+t$. The quasiperiodic lifts cover the full real label line, so the common exchange rule must hold for arbitrary values of the labels. In the rational $SU(2)$ family, the Yang--Baxter equation then requires the three-particle arguments to satisfy
\begin{equation}
 w_{13}=w_{12}+w_{23}.
\label{eq:cocycle}
\end{equation}
The regular solutions are differences of one-particle functions,
\begin{equation}
 w_{RR}=\phi_R(\xi_1)-\phi_R(\xi_2), \qquad w_{RL}=\phi_R(\xi_R)-\phi_L(\xi_L),
\label{eq:cocycleSol}
\end{equation}
with the analogous expression for $w_{LL}$. Redefining the characteristic labels by $\widetilde{\xi}_R=\phi_R(\xi_R)$ and $\widetilde{\xi}_L=\phi_L(\xi_L)$ brings these expressions back to the difference form used above. On the circle, the redefined labels must preserve the common shift generated by one spatial circuit. Thus, within the rational $SU(2)$ family, difference form follows from Yang--Baxter consistency up to this allowed relabeling of the characteristic coordinates.

The local contact prescription specifies how the singular interaction is related to the physical exchange matrix. For one right- and one left-mover, let $r=x_L-x_R$. The incoming side is $r>0$, and the outgoing side is $r<0$. With the symmetric prescription
\begin{equation}
 \delta(r)\Psi(r)  =\frac{1}{2}\delta(r)\bigl[\Psi(0^+)+\Psi(0^-)\bigr],
\label{eq:symmetric}
\end{equation}
the two-body contact operator
\begin{equation}
 h=2i\partial_r+2\delta(r)V
\end{equation}
gives
\begin{equation}
 \left(\id-\frac{i}{2}V\right)\Psi(0^+) =\left(\id+\frac{i}{2}V\right)\Psi(0^-).
\label{eq:jump}
\end{equation}
Defining $\Psi(0^-)=S_{\rm phys}\Psi(0^+)$ gives
\begin{equation}
\begin{aligned}
 S_{\rm phys} &=\left(\id+\frac{i}{2}V\right)^{-1}   \left(\id-\frac{i}{2}V\right),\\
 V &=-2i(\id-S_{\rm phys})(\id+S_{\rm phys})^{-1},
\end{aligned}
\label{eq:Cayley}
\end{equation}
whenever $-1\notin\Spec S_{\rm phys}$.

The symmetric prescription in Eq.~\eqref{eq:symmetric} is part of the model definition. A first-order kinetic operator leaves the product $\delta(r)\Psi(r)$ ambiguous when the wavefunction jumps at the contact~\cite{SutherlandMattis,AlbeverioBook}. We use this prescription throughout, as in Ref.~\cite{Kattel2026}. Within this finite Cayley chart, Hermitian contact operators correspond uniquely to unitary scattering matrices whose spectrum does not contain $-1$. Thus, the scattering matrix fixes the finite contact operator and the coupling used below.

For right-left scattering, $u=\xi_L-\xi_R$ and the physical orientation uses $S_{\rm phys}=S(-u)$. Hence
\begin{equation}
 S_{\rm phys}(u) =\Pi_++\frac{u+i\eta}{u-i\eta}\Pi_- .
\label{eq:Sphys}
\end{equation}
The symmetric channel is free, while the antisymmetric channel gives
\begin{equation}
 V(u)=-\frac{2\eta}{u}\Pi_- =\frac{\eta}{u}(P-\id).
\label{eq:V}
\end{equation}
At $u=0$, the scattering matrix is completely regular and equals $P$. Its antisymmetric eigenvalue is $-1$, which represents a well-defined phase shift with $S=e^{2i\delta}$ and $\delta=\pi/2$ modulo $\pi$. The divergence of $V(u)$ comes from the Cayley parametrization; the scattering matrix and exchange relations remain regular.

The corresponding Hamiltonian is
\begin{equation}
 H_{\rm ch}(t)=H_0+H_{\rm int}(t),
\end{equation}
with
\begin{equation}
 H_0=-i\int\dd x\,\psi_R^\dagger\partial_x\psi_R +i\int\dd x\,\psi_L^\dagger\partial_x\psi_L
\end{equation}
and
\begin{equation}
\begin{aligned}
 H_{\rm int}(t)&=2\int\dd x\,g(x,t)\mathcal O_{RL}(x),\\
 \mathcal O_{RL}(x)&= \NO{\psi_{R,a}^\dagger\psi_{L,b}^\dagger (P-\id)^{ab}{}_{cd}\psi_{L,d}\psi_{R,c}} .
\end{aligned}
\label{eq:Hint}
\end{equation}
In the two-particle sector, this reduces to $2\delta(r)V$ with $V=g(P-\id)$, which fixes the normalization in Eq.~\eqref{eq:Hint} and agrees with Eq.~\eqref{eq:jump}. Comparing with Eq.~\eqref{eq:V} fixes the model-dependent function in Eq.~\eqref{eq:upsilon}:
\begin{equation}
 \Upsilon(g)=\frac{\eta}{g},\qquad u=\frac{\eta}{g(x,t)} .
\label{eq:UpsilonGN}
\end{equation}
Consequently, the abstract condition $\Boxop\Upsilon(g)=0$ becomes
\begin{equation}
 g(x,t)=\frac{\eta}{f_L(x+t)-f_R(x-t)},\qquad \Boxop\left(\frac{1}{g}\right)=0 .
\label{eq:SU2g}
\end{equation}
Using $P=\frac12\id+2t^A\otimes t^A$, where $t^A=\sigma^A/2$ are the $SU(2)$ generators with $\Tr(t^At^B)=\frac12\delta^{AB}$, the interaction can equivalently be written as
\begin{equation}
 \mathcal H_{\rm int}(x,t) =g(x,t)\bigl(4\bm J_R\cdot\bm J_L-n_Rn_L\bigr),
\label{eq:current}
\end{equation}
where $J_\chi^A=\NO{\psi_\chi^\dagger t^A\psi_\chi}$ and $n_\chi=\NO{\psi_\chi^\dagger\psi_\chi}$.

On the circle, the general lifts have the form
\begin{equation}
 f_\chi(s)=\frac{\varpi}{\ell}s+h_\chi(s),\qquad h_\chi(s+\ell)=h_\chi(s).
\label{eq:lifts}
\end{equation}
For example,
\begin{equation}
\begin{aligned}
 f_R(s)&=\kappa s+A\sin\frac{2\pi s}{\ell},\\
 f_L(s)&=\kappa s+B\sin\frac{2\pi s}{\ell}+\beta ,
\end{aligned}
\label{eq:profiles}
\end{equation}
with $\kappa>(2\pi/\ell)\max(|A|,|B|)$ gives $f_R',f_L'>0$ and
\begin{equation}
 g(x,t)= \frac{\eta}{ 2\kappa t+\beta +B\sin\frac{2\pi(x+t)}{\ell} -A\sin\frac{2\pi(x-t)}{\ell}} .
\label{eq:example}
\end{equation}
This coupling is periodic in space and depends nontrivially on both space and time. Its qKZ step is $\varpi=\kappa\ell$. Equal-slope affine maps recover the spatially homogeneous, time-dependent case. The coupling is finite away from the zero set of its denominator, while the exchange matrix remains regular at $u=0$.

For this example, it is useful to separate the spatially uniform part of the scattering variable from its spatial modulation. Write
\begin{align}
 u(x,t)&=2\kappa t+\beta+q(x,t),\\
 q(x,t)&=B\sin\frac{2\pi(x+t)}{\ell} -A\sin\frac{2\pi(x-t)}{\ell} .
\label{eq:qmodulation}
\end{align}
At a fixed time, $q(x,t)$ is a sinusoidal function of $x$. Its maximum absolute value is
\begin{equation}
 \rho(t)=\sqrt{A^2+B^2-2AB\cos\left(\frac{4\pi t}{\ell}\right)} .
\label{eq:rho}
\end{equation}
Thus, the spatial modulation ranges between $-\rho(t)$ and $\rho(t)$. The denominator of the coupling is nonzero at every spatial point at time $t$ precisely when
\begin{equation}
 |2\kappa t+\beta|>\rho(t).
\label{eq:regularslice}
\end{equation}
The quantity $\rho(t)$ is the largest spatial modulation of the denominator. It identifies the time slices on which the finite Cayley coupling is regular throughout the circle. When Eq.~\eqref{eq:regularslice} fails, the denominator vanishes at one or more spatial points. The scattering matrix remains regular at these points because $S(0)=P$; the pole belongs to the chosen finite contact chart, while the factorized transport data remain well defined. On the circle, a zero occurs on a given time slice precisely when the spatial range of the denominator contains zero, namely when Eq.~\eqref{eq:regularslice} fails. Quasiperiodicity does not force a zero on any given slice, but it does force one at some time: at fixed $x$, $u$ grows without bound as $t\to\infty$ and decreases without bound as $t\to-\infty$, so it vanishes in between. On the line, disjoint label ranges can instead keep $g=\eta/u$ finite everywhere. The simpler condition $|2\kappa t+\beta|>|A|+|B|$ is sufficient and shows that the finite Hamiltonian is certainly regular outside a bounded time window.

This construction extends beyond the rational $SU(2)$ example. The same procedure applies to any regular unitary difference-form two-body scattering matrix satisfying braiding unitarity and the Yang--Baxter equation. The characteristic labels provide the transport variables, the two-body matrices define exchange transport, the Yang--Baxter equation gives local path independence, and periodic winding produces compatible qKZ equations. The $SU(2)$ realization shows how one particular scattering matrix can be related, through a chosen contact prescription, to an explicit chiral four-fermion Hamiltonian with a space-time-dependent coupling.

\section{Exact solution of the transport equations}\label{sec:solution}

We use auxiliary variables to construct the spin part of the wavefunction, following the algebraic Bethe ansatz. The auxiliary values are summed over a lattice with spacing $\varpi$. The weight assigned to each lattice point is selected so that the extra terms produced by the transfer matrix cancel between neighboring points. The resulting Jackson sum solves the winding equations. For two particles the sum can be evaluated in closed form and gives the phase accumulated through the sequence of earlier scattering processes. The same construction applies for arbitrary particle number~\cite{Matsuo1993,ReshetikhinJackson1992,TarasovVarchenko1994,BabujianKarowskiZapletal1997}.

The transfer matrix acts on an off-shell Bethe vector in two parts. One part is proportional to the original vector and keeps the same auxiliary roots. The other part replaces one root $\zeta_m$ by the spectral parameter of the transfer matrix. The ordinary Bethe ansatz makes the coefficients of the second part vanish by imposing equations on the auxiliary roots. The present construction uses a different organization of the same algebra. The auxiliary roots are summed over a lattice, and the second part at one lattice point cancels the corresponding contribution at a neighboring point. The explicit one-root calculation is given in Appendix~\ref{app:proof}.

\subsection{Symmetry reduction}

The scattering preserves total spin. We build the highest-weight state of each total-spin multiplet and obtain the rest by applying the total-spin lowering operator $\mathcal S^-=\sum_j\sigma_j^-$. We also use $\mathcal S^+=\sum_j\sigma_j^+$ and $\mathcal S^z=\frac12\sum_j\sigma_j^z$. If $M$ spins are flipped relative to the all-up state, then $\mathcal S^z=N/2-M$. For $M=0$ all spins are aligned and $P-\id$ vanishes. In that sector $F=\qconst(\bm z)\Up$.

\subsection{Off-shell Bethe vectors}

To describe $M$ spin flips, introduce an auxiliary two-state space labelled by $0$. The matrix
\begin{equation}
R_{0k}(v)=v\,\id+i\eta P_{0k}
\label{eq:Raux}
\end{equation}
compares the auxiliary spin with the spin of particle $k$. The spectral parameter $v$ is auxiliary; it is not a physical momentum. The order of the product is part of the definition:
\begin{equation}
\mathbb T_0(v)=R_{0N}(v-\xi_N)\cdots R_{01}(v-\xi_1)
=\begin{pmatrix}\sA(v)&\sB(v)\\ \sC(v)&\sD(v)\end{pmatrix}_0 .
\label{eq:monodromy}
\end{equation}
The entries of $\mathbb T_0$ are operators on the physical spin space. On the all-up vector,
\begin{equation}
\Up=|\uparrow\rangle_1\otimes\cdots\otimes|\uparrow\rangle_N,
\label{eq:vacuum}
\end{equation}
one has $\sA(v)\Up=a(v)\Up$ and $\sD(v)\Up=d(v)\Up$ with
\begin{equation}
a(v)=\prod_{l=1}^N(v-\xi_l+i\eta),\qquad d(v)=\prod_{l=1}^N(v-\xi_l).
\label{eq:ad}
\end{equation}
In the auxiliary basis $(|\uparrow\rangle_0,|\downarrow\rangle_0)$, $\sB(v)=\mathbb T_{12}(v)$ takes the auxiliary spin from down to up. Each permutation $P_{0k}$ preserves the total number of down spins, so this transition moves one down spin from the auxiliary space into the physical chain. Choosing the site $k$ at which the auxiliary line changes channel gives
\begin{equation}
\sB(v)\Up=\sum_{k=1}^N\varphi_k(v)\sigma_k^-\Up ,
\label{eq:Bone}
\end{equation}
where
\begin{equation}
\varphi_k(v)=i\eta\prod_{l<k}(v-\xi_l)
\prod_{l>k}(v-\xi_l+i\eta).
\label{eq:phik}
\end{equation}
The factors containing $+i\eta$ belong to the sites $l>k$, which the auxiliary line meets after the flip, because the monodromy is ordered as $R_{0N}\cdots R_{01}$. For two particles,
\begin{equation}
\sB(v)\Up=i\eta\big[(v-\xi_2+i\eta)\sigma_1^-
+(v-\xi_1)\sigma_2^-\big]\Up .
\label{eq:BoneNtwo}
\end{equation}

The Yang--Baxter algebra gives $[\sB(v),\sB(w)]=0$. An $M$-spin state is obtained by applying these operators at auxiliary values $\zeta_1,\ldots,\zeta_M$. The auxiliary values are unrestricted, giving the off-shell Bethe state
\begin{equation}
\begin{aligned}
\Omega(\bm\xi;\bm\zeta)&=\sB(\zeta_1)\cdots\sB(\zeta_M)\Up\\
&=\sum_{k_1<\cdots<k_M}\sum_{\sigma\in S_M}
\prod_{a<b}\frac{\zeta_{\sigma_a}-\zeta_{\sigma_b}-i\eta}{\zeta_{\sigma_a}-\zeta_{\sigma_b}}\\
&\qquad\times\prod_{a=1}^M\varphi_{k_a}(\zeta_{\sigma_a})\,
\sigma^-_{k_1}\cdots\sigma^-_{k_M}\Up .
\end{aligned}
\label{eq:offshell}
\end{equation}
A second $\sB$ operator does not simply act through Eq.~\eqref{eq:Bone} again. On its way through the chain, its auxiliary line can exchange spins with a site that is already flipped. Summing over these paths gives Eq.~\eqref{eq:offshell}. Each site carries at most one down spin, so the terms have distinct sites $k_1<\cdots<k_M$. The permutation sum assigns the auxiliary values to these sites in all possible orders, and the rational factors are the net effect of the exchanges with already-flipped sites~\cite{KorepinBook}. This is the spin part of the wavefunction. 

\subsection{Summing over the auxiliary values}

We now form the reference amplitude as a weighted sum of the off-shell Bethe vectors $\Omega(\bm\xi;\bm\zeta)$, with a scalar weight $\Phi(\bm\xi,\bm\zeta)$ to be determined. The winding operator produces the wanted term together with terms proportional to $a(\zeta_m)$ and $d(\zeta_m)$, as shown in Appendix~\ref{app:proof}. The lattice sum is designed so that each such contribution is paired with the contribution at the neighboring lattice point. Sums of this type are called Jackson integrals~\cite{Matsuo1993,ReshetikhinJackson1992,TarasovVarchenko1994}.

For one spin flip, the cancellation can be seen directly from the unwanted term in the off-shell action. On the lattice $\gamma=\zeta^0+\varpi\mathbb Z$, the $\sigma_1^-$ component of $F(\bm\xi-\varpi e_1)-\mathcal K_1F(\bm\xi)$ equals $-i\eta$ times
\begin{equation}
\sum_{\zeta\in\gamma}\Phi(\bm\xi,\zeta)\frac{a(\zeta)}{\xi_1-\zeta-\varpi}-\sum_{\zeta\in\gamma}\Phi(\bm\xi,\zeta)\frac{d(\zeta)}{\xi_1-\zeta}.
\label{eq:one-root-unwanted}
\end{equation}
The first sum contains $a(\zeta)$ at one lattice point. To compare it with the second sum, shift that point to its neighbor $\zeta+\varpi$. The denominator then becomes
\begin{equation}
\xi_1-\zeta-\varpi=\xi_1-(\zeta+\varpi).
\end{equation}
The numerators will match after this shift if the one-root weight obeys
\begin{equation}
\Phi(\bm\xi,\zeta)a(\zeta)=\Phi(\bm\xi,\zeta+\varpi)d(\zeta+\varpi).
\label{eq:keyid}
\end{equation}
Inserting Eq.~\eqref{eq:keyid} into the first sum in Eq.~\eqref{eq:one-root-unwanted} gives
\begin{equation}
\sum_{\zeta\in\gamma}\Phi(\bm\xi,\zeta+\varpi)\frac{d(\zeta+\varpi)}{\xi_1-(\zeta+\varpi)}.
\end{equation}
The change of variable $\zeta'=\zeta+\varpi$ leaves $\gamma$ unchanged, so this expression equals
\begin{equation}
\sum_{\zeta'\in\gamma}\Phi(\bm\xi,\zeta')\frac{d(\zeta')}{\xi_1-\zeta'}.
\end{equation}
This is the second sum in Eq.~\eqref{eq:one-root-unwanted}. The unwanted contribution cancels. The identity in Eq.~\eqref{eq:keyid} is the local relation that makes the lattice sum work. Dividing it by $\Phi(\bm\xi,\zeta)d(\zeta+\varpi)$ gives the first-order difference equation
\begin{equation}
\frac{\Phi(\bm\xi,\zeta+\varpi)}{\Phi(\bm\xi,\zeta)}
=\frac{a(\zeta)}{d(\zeta+\varpi)}
=\prod_{i=1}^N\frac{\zeta-\xi_i+i\eta}{\zeta-\xi_i+\varpi}.
\label{eq:Phi-recurrence}
\end{equation}
Introduce the dimensionless coupling
\begin{equation}
\lambda=\frac{\eta}{\varpi}.
\label{eq:lambda}
\end{equation}
To solve this equation, separate one factor of the product and write $x=\zeta-\xi_i$. We seek a function $\cW(x)$ whose shift ratio supplies the corresponding factor
\begin{equation}
\frac{\cW(x+\varpi)}{\cW(x)}=\frac{x+i\eta}{x+\varpi}.
\label{eq:W-shift}
\end{equation}
With $y=x/\varpi$ and $\lambda=\eta/\varpi$, a solution is
\begin{equation}
\cW(x)=\frac{\Gamma(x/\varpi+i\lambda)}{\Gamma(x/\varpi+1)}.
\label{eq:W-solution}
\end{equation}
Indeed, $x\to x+\varpi$ sends $y$ to $y+1$, and $\Gamma(y+1)=y\Gamma(y)$ gives Eq.~\eqref{eq:W-shift}. A factor with period $\varpi$ may multiply $\cW$ and leaves its shift ratio unchanged. These periodic factors belong to the quasi-constant freedom described in Sec.~\ref{sec:quasiconst}. For several spin flips, there is a second ingredient. The off-shell vector contains rational factors involving differences of auxiliary values, and these factors must also be transported when one auxiliary value is shifted. We introduce a pair weight for two auxiliary roots,
\begin{equation}
\cG(u)=\frac{u}{\varpi}\,\frac{\Gamma(u/\varpi-i\lambda)}{\Gamma(u/\varpi+1+i\lambda)} ,
\label{eq:weights}
\end{equation}
\begin{equation}
\Phi(\bm\xi,\bm\zeta)=\prod_{i=1}^N\prod_{m=1}^M\cW(\zeta_m-\xi_i)
\prod_{1\le m<n\le M}\cG(\zeta_n-\zeta_m).
\label{eq:Phi}
\end{equation}
For $M=1$, the pair product is empty and only the factors $\cW(\zeta-\xi_i)$ occur. For $M\ge2$, the off-shell action contains products of the form
\begin{equation}
\prod_{n\ne m}\frac{\zeta_m-\zeta_n-i\eta}{\zeta_m-\zeta_n}.
\end{equation}
The pair weight $\cG(\zeta_n-\zeta_m)$ supplies the extra ratios required when $\zeta_m$ is shifted by one lattice step. The complete weight is the product in Eq.~\eqref{eq:Phi}. The factor $\cW$ couples an auxiliary value to a particle label, while $\cG$ couples two auxiliary values. To verify the shift of $\cG$, set $y=u/\varpi$ and use $\Gamma(y+1)=y\Gamma(y)$:
\begin{align}
\frac{\cG(u+\varpi)}{\cG(u)}&=\frac{y+1}{y}\frac{\Gamma(y+1-i\lambda)}{\Gamma(y-i\lambda)}\frac{\Gamma(y+1+i\lambda)}{\Gamma(y+2+i\lambda)}\nonumber\\
&=\frac{y-i\lambda}{y}\frac{y+1}{y+1+i\lambda}=\frac{u-i\eta}{u}\frac{u+\varpi}{u+\varpi+i\eta}.
\label{eq:G-shift}
\end{align}
The product of the $\cW$ factors gives the particle-dependent ratio in Eq.~\eqref{eq:Phi-recurrence}. When one auxiliary value is shifted in the multi-root case, the factors involving $\cG$ give the additional ratios involving the other auxiliary values.

The one-root cancellation follows by shifting the lattice variable in Eq.~\eqref{eq:keyid}. For several roots, the same argument is applied to one variable $\zeta_m$ at a time. The factors $\cG$ convert the multi-root unwanted terms into discrete differences on the auxiliary lattice. Their sums cancel between neighboring lattice points. The qKZ equations then follow from the summed off-shell action. The calculation is given in Appendix~\ref{app:proof}.

Let $\gamma$ be either a generic lattice $\bm\zeta^0+\varpi\mathbb Z^M$ or a site-anchored lattice
\begin{equation}
\gamma_{\bm k}=\{\zeta_a=\xi_{k_a}+n_a\varpi,\ n_a\in\mathbb Z_{\ge0}\},
\qquad k_1<\cdots<k_M.
\label{eq:anchoredlattice}
\end{equation}
Define
\begin{equation}
F_\gamma(\bm\xi)=\sum_{\bm\zeta\in\gamma}\Phi(\bm\xi,\bm\zeta)\,\Omega(\bm\xi;\bm\zeta).
\label{eq:jackson}
\end{equation}
For one spin flip and a lattice anchored at $\xi_k$, the general sum in Eq.~\eqref{eq:jackson} has the explicit form
\begin{align}
F^{(k)}_m=i\eta\varpi^{N-1}&\sum_{n=0}^\infty\prod_{l=1}^N\frac{\Gamma(n+d_l+i\lambda)}{\Gamma(n+d_l+1)}\nonumber\\
&\times\prod_{l<m}(n+d_l)\prod_{l>m}(n+d_l+i\lambda),
\label{eq:onemagnon}
\end{align}
where $d_l=(\xi_k-\xi_l)/\varpi$ and $F^{(k)}_m$ denotes the component in which the spin at site $m$ is down. This formula follows by substituting $\zeta=\xi_k+n\varpi$ into the weight in Eq.~\eqref{eq:Phi} and the one-spin-flip vector in Eq.~\eqref{eq:Bone}. Its convergence is analyzed next.

For every right-moving winding direction $j$, $F_\gamma$ solves Eq.~\eqref{eq:qKZ}. For a generic lattice it also obeys
\begin{equation}
F_\gamma(\ldots,\xi_{a+1},\xi_a,\ldots)=P_{a,a+1}S_{a,a+1}(\xi_a-\xi_{a+1})F_\gamma(\bm\xi);
\label{eq:bosonic}
\end{equation}
it is a highest-weight vector,
\begin{equation}
\mathcal S^+F_\gamma=0,\qquad
\mathcal S^zF_\gamma=(N/2-M)F_\gamma.
\label{eq:highestweight}
\end{equation}

When a lattice is anchored at a particle label $\xi_k$, the factor $1/\Gamma((\zeta-\xi_k)/\varpi+1)$ in $\cW(\zeta-\xi_k)$ vanishes at $\zeta=\xi_k-\varpi,\xi_k-2\varpi,\ldots$, so the terms extending in the opposite lattice direction vanish. A generic-lattice sum depends on its starting point only modulo $\varpi$, and its terms remain finite as the starting point approaches $\xi_k$. The anchored sums are limits of generic ones and also solve Eq.~\eqref{eq:qKZ}. The resulting one-sided sums are convenient for explicit calculations. Their convergence at the physical value of $\eta$ is discussed in the next subsection.

\subsection{Boundary of convergence and a Coulomb-type phase}\label{sec:convergence}

Consider the one-spin-flip sum in Eq.~\eqref{eq:onemagnon}. Its vector-valued summand can be written as $T_n=\Phi(\bm\xi,\xi_k+n\varpi)\,\Omega(\bm\xi;\xi_k+n\varpi)$. Since $\Gamma(n+a)/\Gamma(n+b)\simeq n^{a-b}$ for large $n$, each of the $N$ factors $\cW$ contributes $n^{-1+i\lambda}$, while the one-spin-flip vector contains polynomials of degree $N-1$ and contributes $n^{N-1}$. Hence
\begin{equation}
T_n\simeq C\,n^{-1+iN\lambda}\bigl[1+O(n^{-1})\bigr],\qquad n\to\infty ,
\label{eq:asym}
\end{equation}
with a constant vector $C$.
For real coupling, the terms decrease only like $1/n$ while their phase keeps rotating, so the sum lies on the boundary of convergence. We define the physical answer by first taking $\operatorname{Im}\eta>0$, where the sum converges absolutely, and then approaching the real axis, $\eta\to\eta+i0$. To describe the tail omitted by cutting the sum off at $n=n_{\rm c}$, set
\begin{equation}
\alpha=iN\lambda .
\label{eq:tailalpha}
\end{equation}
Equation~\eqref{eq:asym} then gives
\begin{equation}
T_n=C\,n^{\alpha-1}\bigl[1+O(n^{-1})\bigr].
\label{eq:tailasym}
\end{equation}

For a function $f$ that decreases sufficiently fast, the Euler--Maclaurin formula gives the first two terms of the sum beyond the cutoff. Since the omitted terms begin at $n=n_{\rm c}+1$, the endpoint at $n_{\rm c}$ enters with weight $-1/2$. Thus
\begin{equation}
\sum_{n>n_{\rm c}}f(n)=\int_{n_{\rm c}}^\infty f(x)\,\mathrm{d}x-\frac{1}{2}f(n_{\rm c})+O\!\left(f'(n_{\rm c})\right).
\label{eq:EM}
\end{equation}
For $\operatorname{Im}\eta>0$, one has $\operatorname{Re}\alpha<0$. Applying Eq.~\eqref{eq:EM} to $f(x)=C x^{\alpha-1}$ gives
\begin{equation}
\int_{n_{\rm c}}^\infty Cx^{\alpha-1}\,\mathrm{d}x=-\frac{C\,n_{\rm c}^{\alpha}}{\alpha}.
\label{eq:tailint}
\end{equation}
Since $T_{n_{\rm c}}\simeq Cn_{\rm c}^{\alpha-1}$ and $\alpha=iN\lambda$, the integral in Eq.~\eqref{eq:tailint} becomes
\begin{equation}
-\frac{C\,n_{\rm c}^{\alpha}}{\alpha}=\frac{i\,n_{\rm c}}{N\lambda}\,T_{n_{\rm c}}.
\label{eq:taillead}
\end{equation}
The endpoint term is
\begin{equation}
-\frac{1}{2}f(n_{\rm c})=-\frac{1}{2}T_{n_{\rm c}}.
\label{eq:tailendpoint}
\end{equation}
The omitted tail is 
\begin{equation}
\sum_{n>n_{\rm c}}T_n=\left(\frac{i\,n_{\rm c}}{N\lambda}-\frac{1}{2}\right)T_{n_{\rm c}}+O\!\left(n_{\rm c}^{-1}\right).
\label{eq:tail}
\end{equation}

The error is of order $n_{\rm c}^{-1}$, the same order as $T_{n_{\rm c}}$ itself, because the $O(n^{-1})$ correction in Eq.~\eqref{eq:tailasym} contributes at this order. The finite-interval Euler--Maclaurin formula explains the cutoff dependence of the partial sum. It reads
\begin{equation}
\begin{aligned}
\sum_{n=1}^{n_{\rm c}}f(n)={}&\int_1^{n_{\rm c}}f(x)\,\mathrm{d}x+\frac{f(1)+f(n_{\rm c})}{2}\\
&+\frac{B_2}{2!}\bigl[f'(n_{\rm c})-f'(1)\bigr]+\cdots,
\end{aligned}
\label{eq:EMfinite}
\end{equation}
where $B_2=1/6$. Terms evaluated at the lower endpoint do not depend on $n_{\rm c}$ and can be included in a constant. Applying Eq.~\eqref{eq:EMfinite} to the leading term in Eq.~\eqref{eq:tailasym} gives
\begin{equation}
\sum_{n=0}^{n_{\rm c}}T_n=F_{\rm fin}+\frac{C\,n_{\rm c}^{\alpha}}{\alpha}+\frac{1}{2}C\,n_{\rm c}^{\alpha-1}+O\!\left(n_{\rm c}^{-1}\right),
\label{eq:partialasym}
\end{equation}
where $F_{\rm fin}$ is independent of $n_{\rm c}$. Using $\alpha=iN\lambda$ and $T_{n_{\rm c}}\simeq Cn_{\rm c}^{\alpha-1}$ gives
\begin{equation}
\sum_{n=0}^{n_{\rm c}}T_n=F_{\rm fin}-\frac{i\,n_{\rm c}}{N\lambda}T_{n_{\rm c}}+\frac{1}{2}T_{n_{\rm c}}+O\!\left(n_{\rm c}^{-1}\right).
\label{eq:partialrewrite}
\end{equation}
The cutoff-dependent part of the partial sum is 
\begin{equation}
-\frac{i\,n_{\rm c}}{N\lambda}T_{n_{\rm c}}+\frac{1}{2}T_{n_{\rm c}}.
\label{eq:partialcutoff}
\end{equation}
The correction supplied by the omitted tail is
\begin{equation}
\frac{i\,n_{\rm c}}{N\lambda}T_{n_{\rm c}}-\frac{1}{2}T_{n_{\rm c}}.
\label{eq:tailcutoff}
\end{equation}
Adding the partial sum and the tail gives
\begin{equation}
\sum_{n=0}^{n_{\rm c}}T_n+\left(\frac{i\,n_{\rm c}}{N\lambda}-\frac{1}{2}\right)T_{n_{\rm c}}=F_{\rm fin}+O\!\left(n_{\rm c}^{-1}\right).
\label{eq:cancellation}
\end{equation}
The two cutoff-dependent terms cancel separately. The regulated amplitude is
\begin{equation}
F_{\gamma_k}=\lim_{n_{\rm c}\to\infty}\left[\sum_{n=0}^{n_{\rm c}}T_n+\left(\frac{i\,n_{\rm c}}{N\lambda}-\frac{1}{2}\right)T_{n_{\rm c}}\right].
\label{eq:regsum}
\end{equation}
For $\operatorname{Im}\eta>0$, this is the convergent sum obtained by adding the asymptotic tail to the partial sum. The real-coupling result is defined by continuing this expression to $\eta\to\eta+i0$. The remaining error vanishes as $n_{\rm c}\to\infty$, so this prescription gives a finite answer for real $\eta\ne0$. For more than one spin flip, we use the same continuation from the region where the sums converge absolutely.

Since $g=\eta/u$, the interaction becomes weaker like $1/u$ as the particles separate in characteristic space. Each right-left scattering process multiplies the singlet amplitude by $(u'+i\eta)/(u'-i\eta)\simeq e^{2i\eta/u'}$, where $u'$ is the label at which it occurs, and successive scattering processes of a pair occur at labels spaced by $\varpi$. For the $n$th earlier process $u'\simeq-n\varpi$, so these phases add up to $-\sum_n2\eta/(n\varpi)\simeq-2\lambda\ln n$, which grows without bound in magnitude; for $N=2$ the same logarithm appears in the factor $n^{2i\lambda}$ of Eq.~\eqref{eq:asym}. A logarithmic phase also occurs in Coulomb scattering, where the potential decays as the inverse distance~\cite{Dollard1964}. The added term in Eq.~\eqref{eq:regsum} removes this divergent constant phase while keeping the finite, label-dependent part of the scattering history.

\subsection{Two particles in closed form}

For one right and one left mover in the incoming region $x_R<x_L$, take $(\xi_1,\xi_2)=(\xi_R,\xi_L)$ and $u=\xi_L-\xi_R$. The one-spin-flip sum already defined in Eq.~\eqref{eq:onemagnon} is regulated according to Sec.~\ref{sec:convergence}. For $N=2$ it becomes a hypergeometric series at unit argument, which is summed by Gauss' theorem ${}_2F_1(a,b;c;1)=\Gamma(c)\Gamma(c-a-b)/[\Gamma(c-a)\Gamma(c-b)]$~\cite{Bailey1935}. This gives
\begin{equation}
F_{RL}=i\eta\varpi\,\frac{\Gamma(1+i\lambda)\,\Gamma(-2i\lambda)}{\Gamma(1-i\lambda)}\,
\fr(u)\,\bigl(|{\uparrow\downarrow}\rangle-|{\downarrow\uparrow}\rangle\bigr),
\label{eq:twobody}
\end{equation}
with
\begin{equation}
\fr(u)=\frac{\Gamma(1-u/\varpi+i\lambda)}{\Gamma(1-u/\varpi-i\lambda)} .
\label{eq:frak}
\end{equation}
The overall prefactor depends on the normalization. The physically important part is the ratio $\fr(u)$. Its Gamma-function form can also be understood directly from the winding equation. In the singlet channel, shifting the relative label by one qKZ step must multiply the amplitude by the two-body scattering phase. Thus
\begin{equation}
 \fr(u+\varpi)=\frac{u+i\eta}{u-i\eta}\,\fr(u),
\label{eq:frshift}
\end{equation}
and Eq.~\eqref{eq:frak} solves this difference equation because $\Gamma(y+1)=y\Gamma(y)$. The Gamma-function ratio obtained by summing the off-shell Bethe vectors is a scalar solution of the two-particle qKZ equation, unique up to a periodic factor. The general two-body amplitude is
\begin{equation}
F_{RL}(\bm\xi)=\sum_{s=-1}^{1}\qconst_s(z_R,z_L)|1,s\rangle+\qconst_0'(z_R,z_L)\,\fr(u)\,|0,0\rangle ,
\label{eq:RLgeneral}
\end{equation}
where $|1,s\rangle$ and $|0,0\rangle$ are the triplet and singlet states of the two spins and the coefficients are arbitrary periodic functions. The triplet part is only constrained to be periodic, because $S$ acts as the identity on the triplet. The complete state is
\begin{equation}
\begin{aligned}
|\Psi_{1,1}(t)\rangle&=\int\dd x_R\,\dd x_L\,\Psi^{ab}(x_R,x_L,t)\\
&\quad\times\psi^\dagger_{Ra}(x_R)\psi^\dagger_{Lb}(x_L)|0\rangle ,\\
\Psi(x_R,x_L,t)&=\begin{cases}F_{RL}(\bm\xi), & x_R<x_L,\\ S_{\rm phys}(u)\,F_{RL}(\bm\xi), & x_R>x_L,\end{cases}
\end{aligned}
\label{eq:RLstate}
\end{equation}
with $\xi_R=f_R(x_R-t)$ and $\xi_L=f_L(x_L+t)$.

For real $u$, the numerator and denominator of Eq.~\eqref{eq:frak} are complex conjugates, so $|\fr(u)|=1$. The ratio changes only the phase of the singlet component. Equation~\eqref{eq:frshift} also gives a simple interpretation of this phase. The singlet amplitude in the outgoing region of Eq.~\eqref{eq:RLstate} equals $\fr(u+\varpi)$, the incoming amplitude for the next scattering step for the same pair. Iterating Eq.~\eqref{eq:frshift} $K$ times gives $\fr(u)=\fr(u-K\varpi)\prod_{n=1}^K(u-n\varpi+i\eta)/(u-n\varpi-i\eta)$. Stirling's formula gives $\fr(u-K\varpi)\simeq K^{2i\lambda}$ for large $K$, and hence
\begin{equation}
\fr(u)=\lim_{K\to\infty}K^{2i\lambda}\prod_{n=1}^K
\frac{u-n\varpi+i\eta}{u-n\varpi-i\eta}.
\label{eq:prodrep}
\end{equation}
Each factor in Eq.~\eqref{eq:prodrep} is the singlet scattering phase from an earlier scattering step for the pair, at relative labels $u-\varpi,u-2\varpi,\ldots$. The Gamma-function ratio resums these scattering phases. Their large-$n$ part behaves as $1/n$, producing the logarithmic phase discussed in Sec.~\ref{sec:convergence}; the factor $K^{2i\lambda}$ removes this divergent overall phase. At weak coupling,
\begin{equation}
\ln\fr(u)=2i\lambda\,\psi_0(1-u/\varpi)+O(\lambda^3),
\label{eq:born}
\end{equation}
 obtained by expanding the logarithm of Eq.~\eqref{eq:frak} to first order in $\lambda$. Here $\psi_0=\Gamma'/\Gamma$ is the digamma function and $\gamma_E$ is Euler's constant. Since $\psi_0(1-y)=-\gamma_E+\sum_{n\ge1}[1/n-1/(n-y)]$ and each scattering process contributes $\ln[(u'+i\eta)/(u'-i\eta)]=2i\eta/u'+O(\eta^3)$, Eq.~\eqref{eq:born} is the first-order Born series $\sum_n 2i\eta/(u-n\varpi)$ with the same regularization. It fails near $u\in\varpi\mathbb N$, where an earlier scattering process occurred at $u=0$ and the coupling $g=\eta/u$ diverged, while the exact amplitude stays unimodular there. At late times, $\fr(u)=\sigma(z)\,(u/\varpi)^{2i\lambda}[1+O(\varpi/u)]$, with $z=z_L/z_R=e^{2\pi iu/\varpi}$, $q=e^{\pi\eta/\varpi}$, and $\sigma(z)=(zq^{-1}-q)/(zq-q^{-1})$; this follows from the reflection formula $\Gamma(y)\Gamma(1-y)=\pi/\sin(\pi y)$ and Stirling's formula. Besides the Coulomb-type phase, the complete scattering history thus leaves a unimodular periodic factor; the same $q$ reappears in the equal-chirality sector below. 

For two particles moving in the same direction, fermionic exchange adds one more factor. The singlet part must be multiplied by a periodic function so that exchanging the particles gives the correct minus sign. This gives
\begin{equation}
\begin{aligned}
F_{RR}(\bm\xi)={}&\sum_{s=-1}^{1}\mathcal A_s(z_1,z_2)|1,s\rangle\\
&+\fr_{RR}(\xi_1-\xi_2)\,\bigl(z_1q^{-1}-z_2q\bigr)\,\mathcal A_0(z_1,z_2)\,|0,0\rangle ,
\end{aligned}
\label{eq:RR}
\end{equation}
where $\fr_{RR}(\Delta)=\Gamma(1+\Delta/\varpi+i\lambda)/\Gamma(1+\Delta/\varpi-i\lambda)$ with $\Delta=\xi_1-\xi_2$, and the functions $\mathcal A$ are periodic and antisymmetric. The function $\fr_{RR}$ alone solves the winding equation in Eq.~\eqref{eq:qKZ}. Under exchange, however, $\fr_{RR}(-\Delta)/\fr_{RR}(\Delta)$ equals the singlet eigenvalue $(\Delta-i\eta)/(\Delta+i\eta)$ of $S(\Delta)$ times $\sigma(z_1/z_2)$, whereas Eq.~\eqref{eq:stat} requires the singlet eigenvalue alone. The factor $z_1q^{-1}-z_2q$, together with an antisymmetric $\mathcal A_0$, changes by exactly $1/\sigma(z_1/z_2)$ under exchange and removes this mismatch. The analytic choice $\mathcal A\propto z_1-z_2$ vanishes when the two right movers coincide on the ring, $\xi_1-\xi_2\in\varpi\mathbb Z$; piecewise-continuous antisymmetric coefficients are also admissible.

\subsection{One magnon: hypergeometric amplitudes}\label{sec:onemagnon}

The one-spin-flip sum was written explicitly in Eq.~\eqref{eq:onemagnon}. We now rewrite its components as generalized hypergeometric functions evaluated at $1$. Each component can be written as
\begin{equation*}
{}_pF_{p-1}\!\left(\left.\begin{matrix}\bm\alpha\\ \bm\beta\end{matrix}\,\right|1\right)=\sum_{n\ge0}\frac{(\alpha_1)_n\cdots(\alpha_p)_n}{n!\,(\beta_1)_n\cdots(\beta_{p-1})_n},
\end{equation*}
with the Pochhammer symbol $(a)_n=\Gamma(a+n)/\Gamma(a)$. For example,
\begin{equation}
\begin{aligned}
F^{(k)}_k&=i\eta\varpi^{N-1}C_k\;
{}_NF_{N-1}\!\left(\left.\begin{matrix}\bm\alpha_k\\ \bm\beta_k\end{matrix}\,\right|1\right),\\
\bm\alpha_k&=\bigl(i\lambda,\,\{d_l+i\lambda\}_{l<k},\,\{d_l+i\lambda+1\}_{l>k}\bigr),\\
\bm\beta_k&=\bigl(\{d_l\}_{l<k},\,\{d_l+1\}_{l>k}\bigr),
\end{aligned}
\label{eq:hyper}
\end{equation}
with
\begin{equation}
C_k=\Gamma(i\lambda)\prod_{l<k}\frac{d_l\,\Gamma(d_l+i\lambda)}{\Gamma(d_l+1)}
\prod_{l>k}\frac{(d_l+i\lambda)\,\Gamma(d_l+i\lambda)}{\Gamma(d_l+1)} .
\end{equation}
The components $m\ne k$ are of the same type with shifted parameters. The parameter excess $\sum_i\beta_i-\sum_i\alpha_i$ of these series equals $-iN\lambda$. A series ${}_pF_{p-1}$ at unit argument converges absolutely only when the real part of its excess is positive, so the physical amplitudes lie on the boundary $\operatorname{Re}(iN\lambda)=0$ of the convergence region and are given by Eq.~\eqref{eq:regsum}. For $N=2$, Eq.~\eqref{eq:hyper} reduces to Gauss' series and reproduces Eq.~\eqref{eq:twobody}. For $N=3$ it gives, for instance, the spin-$1/2$ amplitudes of the sectors $(N_R,N_L)=(1,2)$ and $(2,1)$ as ${}_3F_2$ functions at unit argument.

\subsection{Fermionic statistics and the complete wavefunction}

The generic-lattice sums obey the exchange rule in Eq.~\eqref{eq:bosonic}. Fermionic statistics is then imposed by choosing the periodic coefficients to be antisymmetric within each chirality block. These lattice sums and their descendants provide the solution family used below, with the periodic coefficients supplying the remaining free data~\cite{TarasovVarchenko1997,MukhinVarchenko1997}. The general reference amplitude is
\begin{equation}
F(\bm\xi)=\sum_{r\ge0}\sum_\gamma(\mathcal S^-)^r\,
\qconst_{\gamma,r}(\bm z)\,F_\gamma(\bm\xi).
\label{eq:general}
\end{equation}
Here $\gamma$ runs over generic lattices. The coefficients can contain, for example, the antisymmetric factor $\prod_{a<b\le N_R}(z_a-z_b)\prod_{N_R<a<b}(z_a-z_b)$. They encode the freely chosen incoming chiral wave. The formula below gives the complete highest-weight component explicitly:
\begin{widetext}
\begin{equation}
\begin{aligned}
F_\gamma(\bm\xi)=\sum_{\bm\zeta\in\gamma}\;
&\prod_{i=1}^N\prod_{m=1}^M\frac{\Gamma\bigl((\zeta_m-\xi_i)/\varpi+i\lambda\bigr)}{\Gamma\bigl((\zeta_m-\xi_i)/\varpi+1\bigr)}
\prod_{1\le m<n\le M}\frac{\zeta_n-\zeta_m}{\varpi}\,
\frac{\Gamma\bigl((\zeta_n-\zeta_m)/\varpi-i\lambda\bigr)}{\Gamma\bigl((\zeta_n-\zeta_m)/\varpi+1+i\lambda\bigr)}\\
&\times\sum_{k_1<\cdots<k_M}\sum_{\sigma\in S_M}
\prod_{a<b}\frac{\zeta_{\sigma_a}-\zeta_{\sigma_b}-i\eta}{\zeta_{\sigma_a}-\zeta_{\sigma_b}}
\prod_{a=1}^M\Bigl[i\eta\prod_{l<k_a}(\zeta_{\sigma_a}-\xi_l)\prod_{l>k_a}(\zeta_{\sigma_a}-\xi_l+i\eta)\Bigr]\,
\sigma^-_{k_1}\cdots\sigma^-_{k_M}\Up .
\end{aligned}
\label{eq:fullF}
\end{equation}
\end{widetext}
Finally, insert this reference amplitude into the exchange formula for each coordinate-ordering sector. This gives the exact many-body wavefunction in every fixed-particle-number and total-spin sector. The space-time dependence enters through the characteristic labels $\xi_j$ and the free chiral coordinates $z_j$; the coupling is already encoded in those labels.

\section{Physical consequences}\label{sec:physics}

\subsection{Late-time factorization and a spin-dependent Coulomb phase}

We now consider the many-body form of the logarithmic phase discussed in Sec.~\ref{sec:convergence}. Winding all right movers once around the ring gives
\begin{equation}
\begin{aligned}
\mathcal K^{\rm cl}_R(\bm\xi)&=\mathcal K_{N_R}\Bigl(\bm\xi-\varpi\textstyle\sum_{r<N_R}e_r\Bigr)\cdots\mathcal K_2(\bm\xi-\varpi e_1)\mathcal K_1(\bm\xi)\\
&=\mathfrak C_1\mathfrak C_2\cdots\mathfrak C_{N_R},\\
\mathfrak C_r&=S_{rN}(\Delta_{rN})\cdots S_{r,N_R+1}(\Delta_{r,N_R+1}).
\end{aligned}
\label{eq:cluster}
\end{equation}
All equal-chirality factors cancel by the Yang--Baxter equation and braiding unitarity, so that only right-left matrices remain. Consider the regime in which all right-left label differences are large compared with $\varpi$ and with the spread of labels within each chirality, and let $u\simeq\xi_L-\xi_R$ denote their common value. Expanding Eq.~\eqref{eq:S},
\begin{equation}
\begin{gathered}
\mathcal K^{\rm cl}_R=\id+\frac{i\eta}{u}\,\CRL+O(u^{-2}),\\
\CRL=\sum_{r\in R,\,l\in L}(\id-P_{rl})=\tfrac12N_RN_L-2\,\cS_R\cdot\cS_L ,
\end{gathered}
\label{eq:Ccharge}
\end{equation}
where $\cS_{R,L}$ are the total spins of the two clusters. Winding the right-moving cluster once lowers every right-moving label by $\varpi$ and raises $u$ by $\varpi$, so Eq.~\eqref{eq:qKZ} gives $F(\bm\xi-\varpi\sum_{r\in R}e_r)=\mathcal K^{\rm cl}_RF(\bm\xi)\simeq(\id+i\eta\CRL/u)F(\bm\xi)$. Treating $u$ as continuous on the scale $\varpi$ and integrating, one finds, to leading order,
\begin{equation}
F(\bm\xi)\simeq\Bigl(\frac{u}{\varpi}\Bigr)^{i\lambda\CRL}F_\infty(\bm\xi)\bigl[1+O(\varpi/u)\bigr],
\label{eq:latetime}
\end{equation}
where $F_\infty$ solves the decoupled equal-chirality transport equations. The operator $\CRL$ commutes with $\cS^2$, $\cS_R^2$, and $\cS_L^2$, and its eigenvalue $\frac12N_RN_L-[S(S+1)-S_R(S_R+1)-S_L(S_L+1)]$, with $S$, $S_R$, and $S_L$ the total-spin quantum numbers of all particles and of the two clusters, is the spin-dependent charge of the logarithmic phase. For one pair it equals $0$ in the triplet and $2$ in the singlet, reproducing the late-time behavior of $\fr(u)$ up to the periodic factor $\sigma(z)$. For the example of Eq.~\eqref{eq:profiles}, $u\simeq2\kappa t$, so that the right- and left-moving clusters become free at late times up to the phase $(2t/\ell)^{i\lambda\CRL}$. The charge of this phase depends on the total spins of the two clusters.

\subsection{Constant-coupling limit and the Bethe ansatz}\label{sec:static}

For equal-slope affine maps, $f_\chi(s)=\kappa s+\beta_\chi$, the coupling is $g=\eta/(2\kappa t+\beta)$ with $\beta=\beta_L-\beta_R$, and the step is $\varpi=\kappa\ell$. As $\kappa\to0$ at fixed $\beta$, the coupling becomes the constant $\eta/\beta$ and the step vanishes. Within each chirality the label differences $\kappa(x_j-x_k)$ vanish, so that the equal-chirality matrices tend to $S(0)=P$, while every right-left matrix tends to $S(\mp\beta)$. The winding operators reduce, with corrections of order $\kappa$, to
\begin{equation}
\begin{gathered}
\mathcal K_j\to\mathcal Z_j=X_{j,j-1}\cdots X_{j1}X_{jN}\cdots X_{j,j+1},\\
X_{jk}=\begin{cases}P_{jk}, & \chi_j=\chi_k,\\ S_{jk}(-\chi_j\beta), & \chi_j\ne\chi_k,\end{cases}
\end{gathered}
\label{eq:ALZ}
\end{equation}
which are the transfer operators of the Bethe-ansatz solution of the chiral Gross--Neveu model with constant coupling~\cite{AndreiLowenstein1979,DestriLowenstein1982}, written in the present contact prescription; other prescriptions change the right-left matrices by a scalar phase. Writing $F=\exp(i\sum_jk_j\xi_j/\kappa)\,A(\bm\xi)$, with momenta $k_j$ and an amplitude $A$ that varies slowly with the labels, whose pullback is a product of plane waves $e^{ik_j(x_j-\chi_jt)}$, Eq.~\eqref{eq:qKZ} becomes, to leading order in $\kappa$,
\begin{equation}
e^{-ik_j\ell}A=\mathcal Z_jA ,
\label{eq:ALeigen}
\end{equation}
the eigenvalue problem of the static Bethe ansatz~\cite{AndreiLowenstein1979}. In this limit the qKZ difference equations reduce to the commuting eigenvalue problem of the constant-coupling theory. For $\kappa>0$ the step is fixed by the geometry of the characteristics, and the qKZ system is a genuine difference deformation of this eigenvalue problem.

\subsection{Semiclassical limit and the inhomogeneous Bethe equations}\label{sec:smallstep}

A complementary limit is $\varpi\to0$ at fixed $\eta$ and fixed label differences, i.e., $\lambda=\eta/\varpi\to\infty$. Stirling's formula, $\ln\Gamma(y)\simeq y\ln y-y$ for large $|y|$, then gives $\Phi\simeq e^{-\mathcal Y/\varpi}$, up to a constant, with the master function
\begin{equation}
\begin{aligned}
\mathcal Y(\bm\xi,\bm\zeta)={}&\sum_{i,m}\bigl[\vartheta(\xi_i-\zeta_m-i\eta)-\vartheta(\xi_i-\zeta_m)\bigr]\\
&+\sum_{m<n}\bigl[\vartheta(\zeta_m-\zeta_n+i\eta)-\vartheta(\zeta_m-\zeta_n-i\eta)\bigr],
\end{aligned}
\label{eq:master}
\end{equation}
where $\vartheta(x)=x\ln x-x$. The Jackson integral localizes on critical points $\bm\zeta^\ast$ of $\mathcal Y$, where $\partial_{\zeta_m}\mathcal Y=0$, and $F\simeq e^{-\mathcal Y(\bm\xi,\bm\zeta^\ast)/\varpi}[\Omega(\bm\xi;\bm\zeta^\ast)+O(\varpi)]$~\cite{TarasovVarchenkoBethe1994,TarasovVarchenkoAsym1996}. With $\zeta_m=\mu_m-i\eta/2$, the critical-point conditions are the Bethe equations of the inhomogeneous XXX chain,
\begin{equation}
\prod_{i=1}^N\frac{\mu_m-\xi_i-i\eta/2}{\mu_m-\xi_i+i\eta/2}
=\prod_{n\ne m}\frac{\mu_m-\mu_n-i\eta}{\mu_m-\mu_n+i\eta},
\label{eq:bethe}
\end{equation}
with the characteristic labels as inhomogeneities. On the on-shell vector $\Omega(\bm\xi;\bm\zeta^\ast)$, Yang's operator $\mathcal K_j(\bm\xi;0)$~\cite{Yang1967} acts with eigenvalue
\begin{equation}
\Lambda_j=\prod_{m=1}^M\frac{\xi_j-\mu_m-i\eta/2}{\xi_j-\mu_m+i\eta/2}=e^{\partial_{\xi_j}\mathcal Y},
\label{eq:Lambda}
\end{equation}
and $F(\bm\xi-\varpi e_j)\simeq\Lambda_jF(\bm\xi)$. Locally in label space, the exact amplitude is an instantaneous Bethe eigenstate whose inhomogeneities are transported along the characteristics.

\section{Conclusions}\label{sec:conclusions}

We have constructed integrable chiral theories with space-time-dependent interactions from factorized two-body scattering matrices. The local Hamiltonian realization remains model dependent because different contact prescriptions represent the same scattering matrix through different couplings. In the present work, the explicit wavefunctions are constructed in finite particle-number sectors above the bare Fock vacuum. The same transport construction can, in principle, be formulated for dressed scattering matrices and a filled or dressed reference state, but this requires the corresponding dressed quasiparticle data, normal-ordering terms, and contact reconstruction. We leave this extension to subsequent work.

The concrete theory constructed here is the chiral right-left fermion theory with the local four-fermion Hamiltonian in Eq.~\eqref{eq:Hint} and the spacetime-dependent coupling in Eq.~\eqref{eq:SU2g}. The space-time dependence enters through the characteristic labels. The Yang--Baxter equation makes the local many-body construction path independent, while the qKZ equations impose global winding consistency on the circle. Equal-slope affine maps recover the spatially homogeneous time-dependent case, whereas nonaffine maps produce genuinely space-dependent interactions.

For this theory we have solved the transport problem in closed form. The reference amplitude in every sector is a Jackson integral of off-shell Bethe vectors with the weight of Eq.~\eqref{eq:Phi}, its quasi-constant coefficients are the free chiral data on the ring, and the initial-value problem is solved by the discrete holonomy of Eq.~\eqref{eq:holonomy}. At the physical point the Jackson integrals lie on the boundary of convergence, a direct consequence of the $1/u$ decay of the coupling, and they are fixed by the prescription $\eta\to\eta+i0$, which removes a logarithmic phase of Coulomb type. The two-body amplitude is the Gamma-function ratio of Eq.~\eqref{eq:frak}, one-magnon amplitudes are generalized hypergeometric functions at unit argument, right- and left-moving clusters decouple at late times up to the phase of Eq.~\eqref{eq:latetime}, the constant-coupling Bethe ansatz is recovered as the zero-step limit, Eq.~\eqref{eq:ALeigen}, and the inhomogeneous Bethe equations emerge in the semiclassical limit. The construction may also be applied to trigonometric scattering matrices of the $U(1)$ Thirring and anisotropic Kondo type, and to open boundaries with boundary qKZ equations. Norms and correlation functions require pairings of Jackson integrals.

\begin{acknowledgments}
This work is supported by the Swiss National Science Foundation under grant number 200020-219400. OpenAI ChatGPT (GPT-5.6 Luna) was used to assist with the organization and language editing of a draft prepared by the author.
\end{acknowledgments}

\appendix

\section{Verification of the qKZ equations}\label{app:proof}

We verify the qKZ equation for $j=1$ and $M=1$ on a generic lattice, and then indicate the extensions. Anchored lattices follow as limits of generic lattices, as explained after Eq.~\eqref{eq:highestweight}.

\paragraph*{Transfer matrix at an inhomogeneity.}
Let $\st(v)=\sA(v)+\sD(v)$. Because $R_{01}(0)=i\eta P_{01}$ and the trace over the auxiliary space converts the permutation into a relabeling, one finds
\begin{equation}
\mathcal K_1(\bm\xi;\varpi)=\frac{\st(\xi_1)}{a(\xi_1)},
\label{eq:K1t}
\end{equation}
with no dependence on $\varpi$. Moreover $d(\xi_1)=0$.

\paragraph*{Off-shell action.}
The commutation relations of the Yang--Baxter algebra give
\begin{equation}
\begin{aligned}
\st(v)\sB(\zeta)\Up={}&\Bigl[\frac{v-\zeta-i\eta}{v-\zeta}a(v)
+\frac{v-\zeta+i\eta}{v-\zeta}d(v)\Bigr]\sB(\zeta)\Up\\
&+\frac{i\eta\,[a(\zeta)-d(\zeta)]}{v-\zeta}\,\sB(v)\Up .
\end{aligned}
\label{eq:offshellaction}
\end{equation}
At $v=\xi_1$ the second term in the bracket vanishes, and $\sB(\xi_1)\Up=a(\xi_1)\sigma_1^-\Up$ because $\varphi_m(\xi_1)=0$ for $m\ge2$ and $\varphi_1(\xi_1)=a(\xi_1)$. Hence
\begin{equation}
\begin{aligned}
\mathcal K_1F=\sum_\zeta\Phi\Bigl[&\frac{\xi_1-\zeta-i\eta}{\xi_1-\zeta}\sB(\zeta)\Up\\
&+\frac{i\eta\,[a(\zeta)-d(\zeta)]}{\xi_1-\zeta}\,\sigma_1^-\Up\Bigr].
\end{aligned}
\end{equation}

\paragraph*{Shifted amplitude.}
Under $\xi_1\to\xi_1-\varpi$, Eq.~\eqref{eq:W-shift} gives $\Phi\to\Phi\,(\xi_1-\zeta-i\eta)/(\xi_1-\zeta-\varpi)$. The factor $\varphi_1$ does not depend on $\xi_1$, while $\varphi_m$ with $m\ge2$ acquires $(\zeta-\xi_1+\varpi)/(\zeta-\xi_1)$. The components $m\ge2$ of $F(\bm\xi-\varpi e_1)$ carry the factor $(\xi_1-\zeta-i\eta)/(\xi_1-\zeta)$ and coincide with those of $\mathcal K_1F$. Using $(\xi_1-\zeta-i\eta)\varphi_1(\zeta)=-i\eta\,a(\zeta)$, the $\sigma_1^-$ component of Eq.~\eqref{eq:qKZ} reduces to
\begin{equation}
\sum_{\zeta}\Phi(\bm\xi,\zeta)\frac{a(\zeta)}{\xi_1-\zeta-\varpi}
=\sum_{\zeta}\Phi(\bm\xi,\zeta)\frac{d(\zeta)}{\xi_1-\zeta}.
\end{equation}
By Eq.~\eqref{eq:keyid}, the left-hand side equals $\sum_\zeta\Phi(\bm\xi,\zeta+\varpi)d(\zeta+\varpi)/[\xi_1-(\zeta+\varpi)]$, which becomes the right-hand side after the shift $\zeta\to\zeta-\varpi$ of the summation variable. A generic lattice is invariant under this shift, and both sums converge absolutely for $\operatorname{Im}\eta>0$. This proves Eq.~\eqref{eq:qKZ} for $j=1$ in the region of absolute convergence and hence, by analytic continuation, at the physical point.

\paragraph*{Exchange and the remaining equations.}
The Yang--Baxter equation implies the intertwining relation $\check R_{a,a+1}(\xi_a-\xi_{a+1})\mathbb T_0(v;\bm\xi)=\mathbb T_0(v;s_a\bm\xi)\check R_{a,a+1}(\xi_a-\xi_{a+1})$ with $\check R=PR$ and $s_a\bm\xi$ the label vector with $\xi_a$ and $\xi_{a+1}$ swapped. Since $P\Up=\Up$, it gives $\Omega(s_a\bm\xi;\bm\zeta)=P_{a,a+1}S_{a,a+1}(\xi_a-\xi_{a+1})\Omega(\bm\xi;\bm\zeta)$. Because $\Phi$ is symmetric in $\bm\xi$ and a generic lattice does not depend on $\bm\xi$, Eq.~\eqref{eq:bosonic} follows. The qKZ equations for $j\ge2$ then follow from the equation for $j=1$ by conjugation with exchanges, as in the standard argument of Ref.~\cite{BabujianKarowskiZapletal1997}.

\paragraph*{Several roots and highest weight.}
For $M\ge2$, the action of $\st(\xi_1)$ on $\Omega$ produces, for each $m$, an unwanted term proportional to
\begin{equation*}
a(\zeta_m)\prod_{n\ne m}\frac{\zeta_m-\zeta_n-i\eta}{\zeta_m-\zeta_n}-d(\zeta_m)\prod_{n\ne m}\frac{\zeta_m-\zeta_n+i\eta}{\zeta_m-\zeta_n}.
\end{equation*}
The shift relation in Eq.~\eqref{eq:G-shift} is precisely what is needed for the weight $\Phi$ to convert each of these terms into a total difference in $\zeta_m$, exactly as in the case $M=1$~\cite{BabujianKarowskiZapletal1997,TarasovVarchenko1994}. The highest-weight property follows in the same way from the action of $\sC$, whose unwanted terms are again total differences~\cite{BabujianKarowskiZapletal1997}.

\bibliography{references}

\end{document}